# Optical and mechanical properties of nanofibrillated cellulose: towards a robust platform for next-generation green technologies

*Claudia D. Simão,*[*,a] *Juan S. Reparaz,*[a] *Markus. R. Wagner,*[a] *Bartlomiej Graczykowski,*[a] *Martin Kreuzer,*[a] *Yasser B. Ruiz-Blanco,*[a,b] *Yamila García,*[a] *Jani-Markus Malho,*[‡c] *Alejandro R. Goñi,*[d,e] *Jouni Ahopelto*[c] *and Clivia M. Sotomayor Torres*[a,e]

a ICN2 Catalan Institute of Nanoscience and Nanotechnology, Campus UAB, 08193 Bellaterra (Barcelona), Spain

b Unit of Computer-Aided Molecular "Biosilico" Discovery and Bioinformatic Research (CAMD-BIR Unit), Faculty of Chemistry-Pharmacy. Universidad Central "Marta Abreu" de Las Villas, Santa Clara, 54830, Villa Clara, Cuba

c VTT Technical Research Centre of Finland, P.O. Box 1000, FI-02044 VTT, Espoo, Finland

d Institut de Ciencia de Materials de Barcelona, ICMAB-CSIC, Barcelona, Spain

e Catalan Institute of Research and Advanced Studies (ICREA), Barcelona 08010, Spain

**Abstract**

Nanofibrillated cellulose, a polymer that can be obtained from one of the most abundant biopolymers in Nature, is being increasingly explored due to its outstanding properties for packaging and device applications. Still, open challenges in engineering its intrinsic properties remain to address. To elucidate the optical and mechanical stability of nanofibrillated cellulose as a standalone platform, herein we report on three main findings: i) for the first time an experimental determination of the optical band gap of nanofibrillated cellulose, important for future modelling purposes, based on the onset of the optical band gap of the nanofibrillated cellulose film at $E_g \approx$ 275 nm (4.5 eV), obtained using absorption and cathodoluminescence measurements. In addition, comparing this result with ab-initio calculations of the electronic structure the exciton binding energy is estimated to be $E_{ex} \approx$ 800 meV; ii) Hydrostatic pressure experiments revealed that nanofibrillated cellulose is structurally stable at least up to 1.2 GPa; iii) Surface elastic properties with repeatability better than 5% were observed under moisture cycles with changes of the Young modulus as large as 65%. The results obtained show the precise determination of significant properties as elastic properties and interactions that are compared with similar works and, moreover, demonstrate that nanofibrillated cellulose properties can be reversibly controlled, supporting the extended potential of nanofibrillated cellulose as a robust platform for green-technology applications.

# 1. Introduction

Reducing the consumption of oil-based materials and their replacement with renewable and recyclable products is one of the most important challenges in today's society. Nanocellulose, a derivative of cellulose, occurs in two forms, cellulose nanocrystals or whiskers (CNC)(Azizi Samir, Alloin, & Dufresne, 2005), and nanofibrillated cellulose (NFC)(Isogai, Saito, & Fukuzumi, 2011). Nanocellulose has gained attention as one of the most promising candidates for sustainable replacement of oil-based plastics(Turbak, Snyder, & Sandberg, 1983)(Pääkkö et al., 2007) not only because of its abundance in nature, but also due to its low density, biodegradability, well-established industrial use, and its low cost (Klemm et al., 2011), since some types of nanocelluloses are difficult to obtain in large scale and required some purification methods making some of these processes expensive; all crucial characteristics for technological applications. This biopolymer can be extracted from plants using mechanical pulping and it can be easily dispersed in water, providing a cleaner route to production as it is a renewable and recyclable material(Oksman, Mathew, & Sain, 2009),(Abdul Khalil et al., 2014). Its properties can vary drastically with the processing method employed(Eichhorn et al., 2009). Nevertheless, the outstanding properties of nanocellulose have captured the attention of a wide range of fields seeking non-hazardous disposable devices with high stability, mainly as a composite material for packaging(Aulin, Salazar-Alvarez, & Lindström, 2012),(Lindström, Karabulut, Kulachenko, Sehaqui, & Wågberg, 2012), electronics(Malho, Laaksonen, Walther, Ikkala, & Linder, 2012), energy storage(Nyholm, Nyström, Mihranyan, & Strømme, 2011), and energy conversion(Liu, Tao, Bai, & Liu, 2012). However, the overall properties of NFC are known to depend strongly on extraction and purification processes.(Lee, Buldum, Mantalaris, & Bismarck, 2013) In particular, one of the major drawbacks shown by NFC is the poor stability with respect to moisture, that affects its mechanical integrity.(Lindström et al., 2012) As a consequence, the ability to tailor its elastic constants is essential to promote NFC as a promising candidate to replace oil-based materials(Malho et al., 2012),(Dufresne, 2013). The application of NFC as a supporting platform for sensing applications has been increasingly reported in recent publications. For example, Nogi et al. demonstrated that NFC films are exceptionally good substrates for optical devices due its remarkable optical transparency.(Nogi, Iwamoto, Nakagaito, & Yano, 2009) Wang et al. showed the advantages of combining NFC with carbon nanotubes in hybrid aerogels to enhance mechanical and electrical properties(Wang et al., 2013), whereas Yan et al. followed a similar approach by mixing NFC with graphene leading to highly stretchable piezoresistive nanopaper sheets(Yan et al., 2014) In spite of the progress observed, the application of cellulose in devices appears restricted to the role of matrix in composite materials and it's applicability is usually limited under non ideal moisture conditions. Here we investigate and determine NFC fundamental optical, mechanical and vibrational properties to extend the application of NFC films as building blocks for nanodevices.

**2. Experimental**

*2.1. Materials.*

The nanofibrillated cellulose (NFC) film was prepared from an aqueous native birch pulp by first grinding and then passing through a homogenizer as already reported elsewhere(Isogai et al., 2011), as depicted in Figure 1a. Preparation of NFC films was carried out using a tip sonicator (Vibra-Cell VCX 750, Sonics & Materials Inc.) to enhance the dispersity of the NFC fibrils. 2.5kJ of energy was applied to the fibril dispersions using 40 % of the full output, because a light sonication is known to open up the fine structure of the NFC aggregates and increase the homogeneity of the dispersion. The fibril dispersions were pipetted in an ultrasound sonicator to remove the bubbles before film formation. Free-standing films were then created by vacuum filtration from aqueous fibril dispersions, wherein NFC concentration was 2.0 g/l. The dispersions were filtrated using a Durapore membrane (GVWP, 0.22μm, Millipore, U.S.A.) and an O-ring to determine the diameter of the film. A gentle press was applied to the films after filtration using a 300 g load for 10 min to prevent wrinkling. At the end the films were dried overnight in an oven at +65 °C.

*2.2. Methods*

*2.2.1. Structural characterization.* For a top view image, the NFC sample was mounted with carbon tape on a sample holder and measured with a FEI Quanta 650 FEG environmental scanning electron microscope using an electron energy of 2 keV. In order to measure the cross section of the NFC film, the sample was torn apart in liquid nitrogen. The cross section was measured using a FEI Magellan 400L extreme resolution scanning electron microscope using 1keV electron energy. X-ray diffraction patterns of the NFC film were obtained by a PANalytical X'Pert PRO diffractometer using a Cu Kα radiation ($\lambda = 1.54$ Å) with a current of 40 mA and an anode voltage of 45 kV. Data were recorded from detector angle $2\theta = 12°$ to $26°$ with a step size of 0.05°. Surface analysis of the sample was performed by X-ray photoelectron spectroscopy (XPS) experiments using a PHI 5500 Multitechnique System (from Physical electronics) with a monochromatic X-ray source (Al K α line of 1486.6eV energy and 350 W), placed perpendicular to the analyzer axis and calibrated using the 3d5/2 line of Ag with a full width at half maximum (FWHM) of 0.8 eV. The analyzed area was a circle of 0.8mm diameter, and the selected resolution for the spectra was 187.5eV of Pass Energy and 0.8 eV/step for the general spectra and 23.5eV of Pass Energy and 0.1 eV/step for the spectra of the different elements. All measurements were made in an ultra-high vacuum chamber pressure between $5 \times 10^{-9}$ and $2 \times 10^{-8}$ torr. Raman scattering spectra were collected using a confocal LabRam HR800 spectrometer (Horiba JobinYvon) with spectral resolution better than 0.5 $cm^{-1}$. Measurements under pressure were carried out using the diamond anvil cell (DAC) technique. A 4:1 mixture of methanol and ethanol was employed as the pressure-transmitting medium. Pressure was monitored in situ by the shift in the ruby R1 line. Fourier Transform Infrared Spectroscopy (FTIR) was performed with a Bruker spectrometer using an attenuated total reflection setup, with a DTGS detector from 4000 $cm^{-1}$ to 400 $cm^{-1}$ with a spectral resolution of 4 $cm^{-1}$ and 32 scans per spectrum.

*2.2.2. Optical characterization.* The absorbance and circular dichroisme spectra of the nanocellulose film were measured on a UV-Vis Spectrophotometer Jasco 715 with a polarimeter power supply PS-150 J at 20ºC. Cathodoluminescence spectra were recorded using a Gatan MonoCL2 grating spectrometer attached to a Philips XL30 scanning electron microscope (SEM). The spectra were acquired with 5kV at 1000x magnification with a sample current of 100nA.

*2.2.3. Thermal characterization.* Dynamic thermogravimetric measurements were performed by using a Mettler TGA/SDTA851e instrument. Temperature programs for dynamic tests were run from 30 ºC to 300 ºC at a heating rate of 10 ºC/min. These tests were carried out under anhydrous nitrogen atmosphere (50 ml/min) in order to prevent any thermoxidative degradation inside the aluminium capsule (70μL). Differential scanning calorimetry were carried out in a DSC822e de Mettler Toledo from room temperature to 300 ºC at a heating rate of 10 ºC/ min under anhydrous nitrogen atmosphere (50 ml/min) in an aluminium capsule (40μL).

*2.2.4. Mechanical characterization.* Nanoindentation experiments were carried out using a commercial a UMIS indenter from Fischer–Cripps Laboratory, operated in the load-control mode and using a Berkovich-type, pyramid-shaped, diamond indenter. Peak Force Quantitative NanoMechanics mode (QNM) atomic force microscopy (AFM) was performed using a Multimode 8 Nanoscope V electronics. Commercially available AFM probe TAP 525 (nominal spring constant: 40nN/nm) from Bruker. Spring constant (measured by thermal noise method) 17: 43,8nN/nm. Sensitivity: 57nm/V. Set point: 250nN. Peak force amplitude: 50nm. Force limit in DMT mode: 2μN. DMT modulus limit: 20GPa. The experiments were performed inside an inert gas chamber with controllable moisture. The humidity was controlled by a flow of inert nitrogen stream for anhydrous environment or a stream of nitrogen bubbled through a sparger containing milipore water for a moist atmosphere. The moisture of the chamber was monitored by an Oregon Scientific humidity sensor. Brillouin spectroscopy measurements were performed on a six-pass tandem type interferometer (JRS Scientific Instruments) using 180° (backscattering) and 90° geometries. The 514.5 nm laser line of an argon-ion laser equipped with an ethalon was used as excitation source. The elastic constant $C_{11}$ and $C_{12}$ of NFCs were determined by recording the Brillouin spectra for longitudinal and transverse acoustic waves. We note that for isotropic materials in backscattering geometry (180) only longitudinal modes are observable due to scattering selection rules.(Vacher & Boyer, 1972) Since the velocities of transverse modes (silent in backscattering geometry) are also necessary to determine the elastic properties, we also used the 90R scattering geometry.

## 3. Results and Discussion

*3.1. Structural and elemental analysis.* Figure 1b displays scanning electron microscopy (SEM) images of the NFC foil. The NFC sample consists of a broad distribution of cellulose fibers with diameters ranging between 20 and 200 nm (see Supporting Information, Figure S1). The cross sectional image shown in the inset reveals a film thickness of about 10 μm. Figure 1c displays the X-ray diffraction (XRD) pattern of the NFC sample, which is typical for monoclinic cellulose I crystallites. In order to extract the crystallite dimensions, we use the (101), (10$\bar{1}$), (021) and (002) reflections following the procedure previously described by Meyer *et al.*(Meyer & Misch, 1937). The large linewidth of all four peaks arises from the small dimensions of the crystallites, as given by β in Table S1 of the Supporting Information section. The average crystallite size was determined to be in the range of 2 to 4 nm which is comparable to previously reported values(Garvey, Parker, & Simon, 2005)'(Cao & Tan, 2005)'(Fink, Hofmann, & Philipp, 1995). The NCL sample showed a diffraction pattern of a highly crystalline cellulose Iβ structure (native cellulose)(Cao & Tan, 2005) with a crystallinity index of 79% (see Supporting Information). The structure can be described by a monoclinic unit cell which contains two parallel aligned cellulose chains, with the presence of an amorphous part, which is in good agreement with the literature(Ruel, Nishiyama, & Joseleau, 2012).

The surface elemental composition of the NFC film was investigated using X-ray photoelectron spectroscopy (XPS) as shown in Figure 1d. The XPS overview spectrum shows the presence of carbon, oxygen and silicon. The onset of the valence band region is attributed to the HOMO level at an energy of $E_{HOMO}$= 3.81 eV with respect to Ag. The high resolution spectra of the C 1s peak revealing the presence of the three carbon bonds types in the sample: C1s-C tertiary (284.19eV), C1s-C sp2 (285.84eV) and at higher binding energies C1s-OH (286.67 eV).

Moreover, at the valence-band region (0 – 40 eV), the O(2s) peak, as well as in the region at binding energies below 18 eV provide analytical information allowing distinctions to be made regarding surface functionality(Xie & Sherwood, 1991). In this region, the O2s state appears at 25.79 eV and the C2s at 18.52 eV, showing a separation of ≈7.28 eV. The region below 18 eV shows a combination of three peaks in contrast with the same type of valence band spectra measured and calculated in β-cellulose and lignin where a doublet is observed(Haensel et al., 2012). These triplet peaks is related to bridged intramolecular hydrogen bonding in the surface region(Xie & Sherwood, 1991).

Simao et al. Carbohydrate Polymers, doi:10.1016/j.carbpol.2015.03.032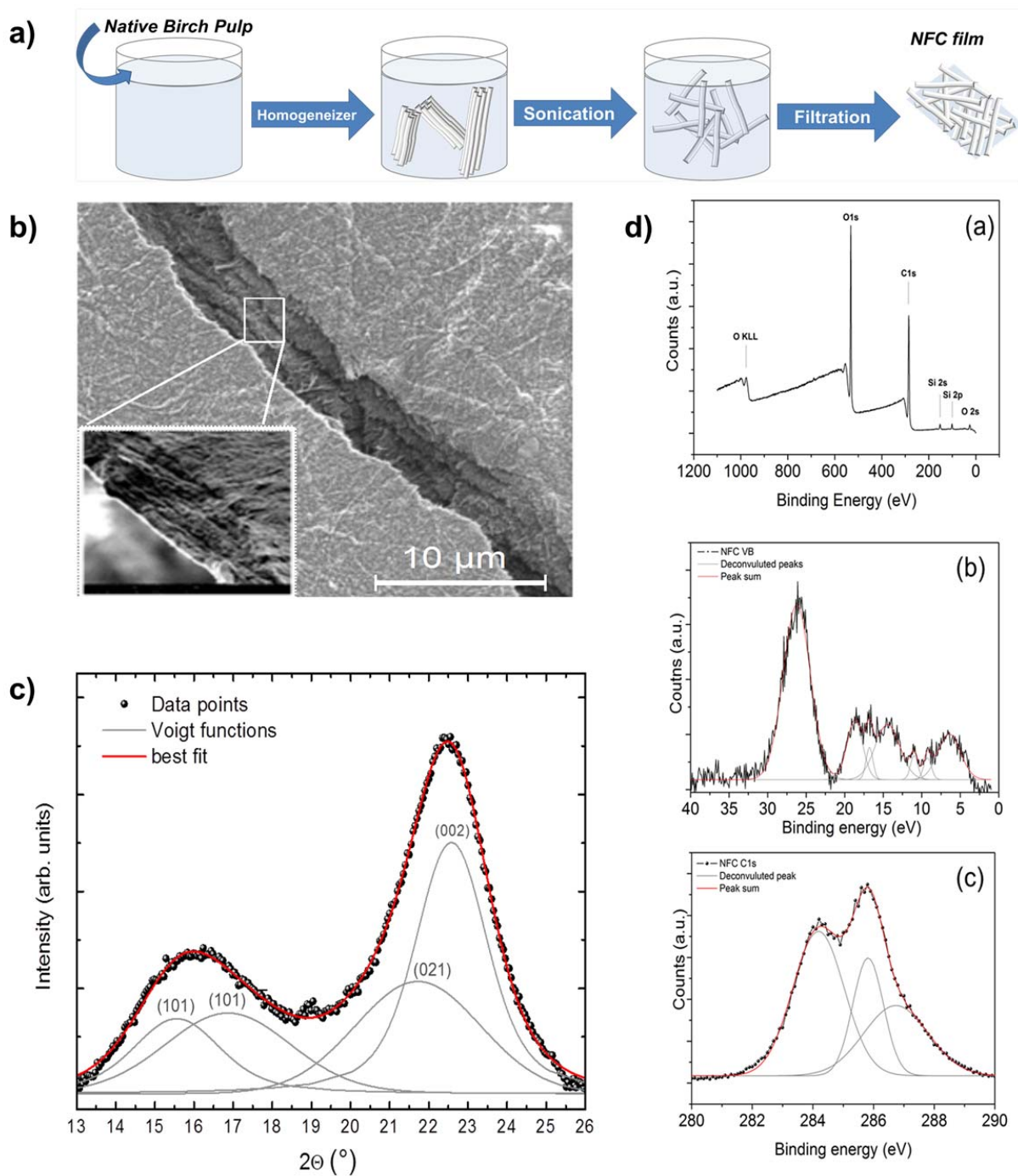

**Figure 1:** a) Schematic representation of the preparation process of the NFC film. b) SEM top view and cross sectional image (inset) of the NFC foil. c) XRD 2-theta scan: Solid symbols represent measured data points, the red line is a fit to the data points using four Voigt functions

(light gray lines). d) X-ray photoelectron spectroscopy general (a), high resolution carbon (C 1s) (b) and valence band region (c) spectra.

*3.2. Thermal behavior.* Thermal degradation patterns and differential scanning calorimetry (DSC) of the NFC film were recorded (Figure S2) and the obtained parameters are summarized in Table S2 of the Supporting Information section. A small weight loss of 3.2 % from 29 to 82 ºC was attributed to the release of moisture from the samples. The onset of a second weight loss step observed at 250 ºC is ascribed to the cellulose decomposition(Morán, Alvarez, Cyras, & Vázquez, 2007). In the DSC spectra is observed a first endothermic phenomenon with an onset at 108ºC with an associated heat of 31.3J/g, a second exothermic phenomenon at 146ºC with an associated heat of 15.6 J/g followed by an exothermic phenomenon with an associated heat of 14.6J/g. Finally, an exothermic phenomenon is registered at 280ºC (associated heat 22.6 J/g, after an endothermic phenomenon with an associated heat of 23.0 J/g.

*3.3. Vibrational properties and pressure stability.* Fourier transform infrared (FTIR) spectrum of the NFC film shows the characteristic absorption peaks of cellulose without presence of impurities (see Supporting Information, Figure S3). Raman spectroscopy revealed comparable results to a previously reported work by Agarwal *et al*. (Agarwal, Reiner, & Ralph, 2010) in which the authors used FT-Raman spectroscopy to determine the crystallinity of nanocellulose samples in diverse forms. Furthermore, we investigated the hydrostatic pressure dependence of the Raman modes using a diamond anvil cell (Figure 2).

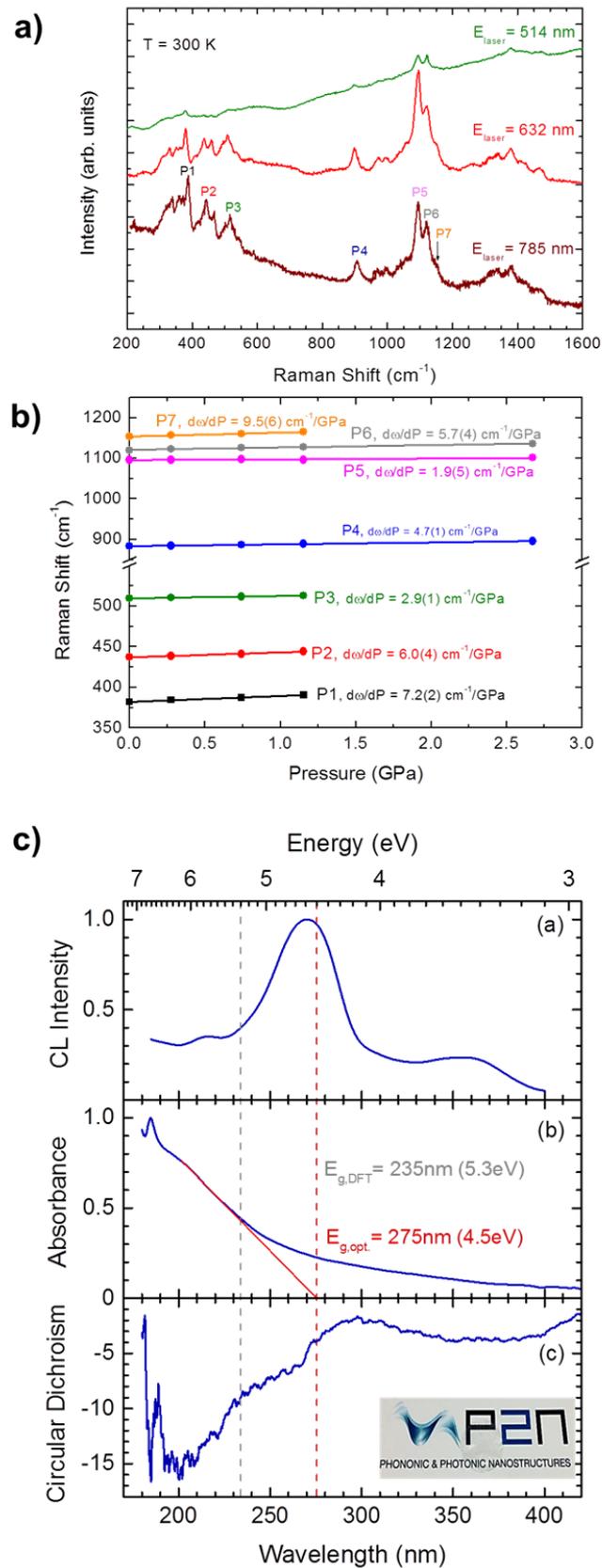

**Figure 2:** a) Raman spectra of NFC film at 300 K using different excitation lines: 514.5, 632, and 785 nm. The variations of the background levels arise from the different absorption coefficients at these three excitation wavelengths. b) Hydrostatic pressure dependence of the labeled Raman modes (P1 to P7) in Figure 2a using the 785 nm excitation line. The numbers in brackets indicate the experimental error of the hydrostatic pressure coefficients. c) Optical characterization of NFC film: (a) Cathodoluminescence, (b) absorption, and (c) circular dichroism spectra of NFC in the UV from 180 to 420 nm. The optical band gap is determined from the absorption onset (solid red line) and indicated by the vertical dashed line at 275 nm. Inset in (c) shows picture of the NFC film prepared.

All observed modes show a linear dependence as function of pressure with hydrostatic pressure coefficients varying between 2 and 10 cm$^{-1}$/GPa. The pressure coefficients and the corresponding Grüneisen parameters were obtained using the bulk modulus determined by Brillouin scattering measurements and are listed in Table 1. To our knowledge, there is no comparable data available in the literature except for biaxial stress measurements of the 1095 cm$^{-1}$ peak.(Agarwal et al., 2010),(Sturcová, Davies, & Eichhorn, 2005),(Hsieh, Yano, Nogi, & Eichhorn, 2008) Our studies show that the NFC film starts to degrade at pressures higher than 1.2 GPa, where the intensity of most Raman peaks gradually vanishes. At 2.8 GPa, the highest pressure attained the Raman signal was no longer detectable. This behavior implies that the hydrostatic pressure in that range induces a non-reversible structural modification of the fibrils. The precise nature of this modification remains subject to future studies. In relation with reported stress sensing capabilities of NFC,(Wang et al., 2013) the present high pressure studies set a limit for higher operation pressure of these devices.

**Table 1.** Pressure coefficients for Raman modes.

|  | **P1** | **P2** | **P3** | **P4** | **P5** | **P6** | **P7** |
|---|---|---|---|---|---|---|---|
| **d$\omega$/dP** [cm$^{-1}$/GPa] | 7.2 ±0.2 | 6.0±0.4 | 2.9±0.1 | 4.7±0.1 | 1.9±0.5 | 5.7±0.4 | 9.5±0.6 |
| **$\omega_0$** [cm$^{-1}$] | 382.0±0.3 | 437.1±0.3 | 509.7±0.3 | 883.2±0.3 | 1031.4±0.4 | 1095.9±0.3 | 1120.0±0.3 |
| $\gamma_i$ | 0.21±0.01 | 0.15±0.01 | 0.06±0.01 | 0.06±0.01 | 0.02±0.01 | 0.06±0.01 | 0.09±0.01 |

*3.4. NFC bandgap and optical properties.* To determine the fundamental bandgap of the NFC foil we performed cathodoluminescence (CL) and absorption measurements. Figures 2c display the CL and absorption spectra of the NFC foil at room temperature. The optical band gap is determined by the onset of the absorption edge at about 275 nm (4.5 eV), which is in the range of most insulators. This wavelength is in excellent agreement with the luminescence maximum at 270 nm in the CL spectra which originates from the radiative recombination of excited charge carriers (Frenkel excitons) between the LUMO and HOMO states. The absorption tail observed at longer wavelengths is attributed to defects within the band gap as demonstrated by the presence of a weak luminescence band around 360 nm and by the excitation-dependent Raman measurements. No pronounced absorption was observed at 260 and 280 nm indicating that the polysaccharides contained no nucleic acids, proteins, or polypeptides(El Rassi & Smith, 1995). The sharp peak at 185 nm in the absorption spectrum is typical for carbohydrates and the exceptional transmittance is translated in high transparency. The circular dichroism (CD) spectrum in Figure 2c shows the chiroptical activity in the same spectral region with a negative absorption at approximately 200 nm followed by a positive absorption at 190 nm. This behavior is characteristic of the beta-sheet conformation, which is in agreement with the collected crystallographic data. We performed *ab-initio* calculations of the electronic structure which resulted in an electronic band gap energy of $E_g$= 5.3 eV (see Supporting Information). Considering that the optical band gap as derived from the onset of the absorption edge is given by the electronic band gap minus the binding energy of the Frenkel excitons, a rough estimate of the exciton binding energy of about 800 meV is obtained.

*3.5. Mechanical stability and elastic constants.* Nanoindentation experiments were performed by mechanically attaching the sample to a silicon substrate. The Young modulus of the NFC is obtained from the typical load versus displacement curves (see Supporting Information, Figure S4) for an average relative humidity of 80% during the measurements and was found to be $E \approx 2$ GPa. This value is in good agreement with previously reported ones at a comparable ambient humidity [25]. Higher values of the Young modulus are expected at lower humidity ranging from 1 to 18 GPa(Rusli & Eichhorn, 2008). Brillouin light scattering (BLS), which probes the longitudinal acoustic (LA) and transverse acoustic (TA) waves provided direct access to the elastic constants of the NFC crystallites ($C_{11}$ and $C_{12}$), the phase velocities of these modes ($v_{LA}$ and $v_{TA}$), as well as the Young modulus (E) and Bulk modulus (B). An advantage of this technique compared to nanoindentation is its large insensitivity to the moisture content since the acoustics modes are mostly not influenced by the presence of water. In addition, the large exposure time to the laser radiation (hours) ensures a water-free local environment. However, for practical applications the value obtained from the nanoindentation experiments is more probably more appropriate since it reflects the mechanical behavior of NFC including moisture.

Representative Brillouin spectra at room temperature are shown in Figure 3 in backscattering (180) and perpendicular (90R) geometries. The phase velocities of both acoustic modes are determined from their frequency shift with respect to the incident laser wavelength and the refractive index of the medium(Speziale et al., 2003). Based on these measurements, we obtain values of $v_{LA}$ = 3427 ± 22 m/s and $v_{TA}$ = 1783 ± 19 m/s for the longitudinal and transversal acoustic phase velocities, respectively. In combination with the mass density $\rho$ of NFC we derive the elastic constants as $C_{11} = \rho v_L^2 = 17.61 \pm 0.23$ GPa and $C_{12} = \rho v_L^2 - 2\rho v_T^2 = 8.07 \pm 0.29$ GPa assuming an isotropic mass density of $\rho = 1500 \, kg/m^3$. Finally, the elastic moduli are obtained by a simple combination of $C_{11}$ and $C_{12}$. All derived elastic parameters are summarized in Table 2.

**Table 2.** Phase velocities of the LA and TA modes, elastic constants $C_{11}$ and $C_{12}$, Poisson ratio µ, Young E and Bulk B moduli for NFC.

| Sample | $v_{LA}$ (m/s) | $v_{TA}$ (m/s) | $C_{11}$ (GPa) | $C_{12}$ (GPa) | µ | E (GPa) | B (GPa) |
|---|---|---|---|---|---|---|---|
| NFC | 3427±22 | 1783±19 | 17.61±0.23 | 8.07±0.29 | 0.31±0.01 | 12.5±0.4 | 11.0±0.5 |

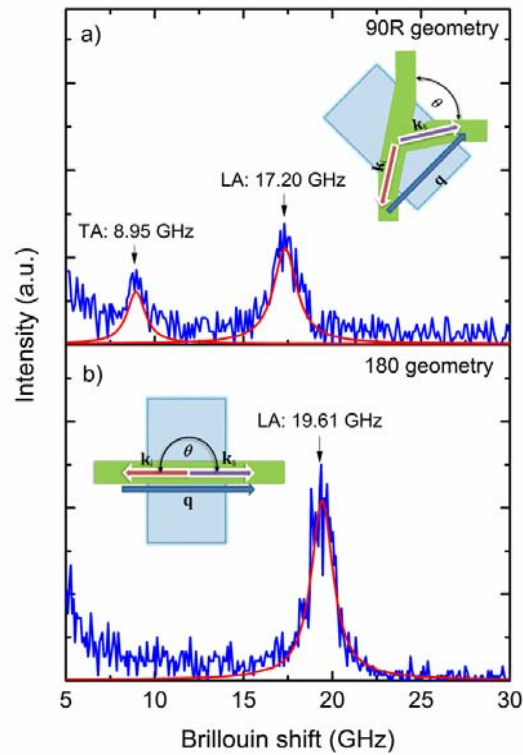

**Figure 3:** Anti-Stokes components of BLS spectra for (a) 90R and (b) 180 scattering geometries. The experimental data were fitted using Lorentzian functions as shown with red full lines. $k_i$ and $k_s$ are the wavevectors of the incident and scattered light, and q is the acoustic phonon wavevector.

Nanoindentation and Brillouin spectroscopy measurements yield values of the elastic moduli of NFC film for two extreme cases: the former is under conditions of a high relative humidity of 80% and the latter is the case equivalent to a water-free environment. However, for a full description of the elastic properties a systematic study as a function of the relative humidity is essential. We carried out measurements using quantitative nanomechanical peak atomic force microscopy (QNM-AFM) in a controlled humidity environment. This study provides 2-dimensional maps of the local surface Young modulus and the adhesion force between the AFM tip and the surface of the NFC sample. Figure 4a displays average roughness and representative topography line scans of the NFC sample under different relative humidity conditions. A flattening of the surface roughness becomes apparent with no changes in the fibrillated structure as the relative humidity is increased. The Young modulus and adhesion force for each relative humidity are shown in Figure 4b and 4c. At a relative humidity of 0%, the Young modulus is 16.4 GPa and the adhesion force is 8 nN. At 100% relative humidity the Young modulus is reduced to 6 GPa and the adhesion force drops abruptly close to zero, which can be attributed to water condensation at the surface of the NFC. Three complete moisture cycles between 0% and 100% were performed exhibiting an extraordinary repeatability better than 95%, i.e., the nano-mechanical properties of NFC are maintained upon moisture cycling. In earlier reports of cellulose based materials with increasing amount of NFC, the Young modulus increases with the increase of the crystallinity of the material (obtained with an higher amount of NFC in the composite) and usually decreases with moist(Hulleman, Kalisvaart, Janssen, Feil, & Vliegenthart, 1999)(Müller, Laurindo, & Yamashita, 2009). In our work, we also observed the decrease of the NFC film superficial Young modulus with increasing relative humidity, thus this behavior indicates that the NFC film loses crystallinity with increasing moist, but on another hand, it is noticeable the non-destructive interaction between moist and the surface of pure NFC film, since the values are reversible and reproducible in different cycles. This result demonstrates an excellent stability in comparison with previous studies of the mechanosorptive behavior of NFC based materials showing that the moisture dependent creep behavior arises from the interfibril bonding(Lindström et al., 2012) and it is worth noticing that the Young modulus obtained for low moisture conditions are in accordance with previous studies(Cranston et al., 2011). The stability against moisture cycling and the ability of tuning the elastic properties paves the way towards a wide field of applications based on this biodegradable and low cost material.

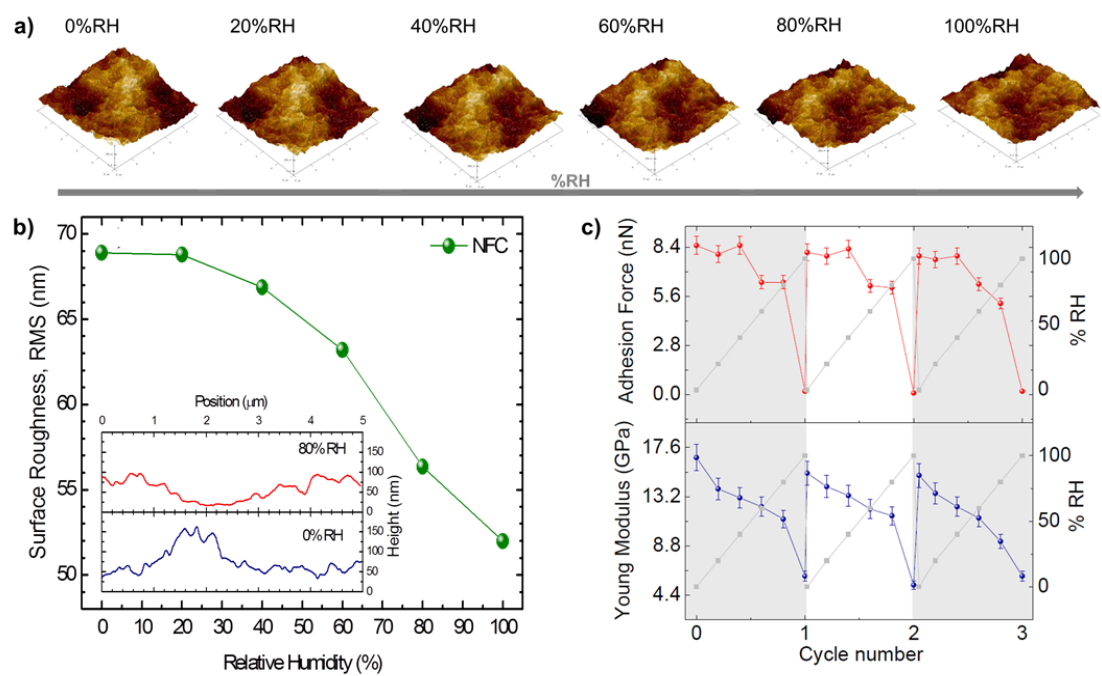

**Figure 4:** AFM Peak Force measurements under different moisture conditions. a) 3D topography QNM AFM images. b) Surface roughness (RMS) as a function of relative humidity (%RH). The inset shows two representative scans at 0 and 80 % RH. c) Adhesion force (top) and Young modulus (bottom) as a function of moisture percentage and cycle number and schematic representation of the NFC foil surface in equilibrium with environmental moisture.

## 4. Conclusions

The nanofibrillated cellulose (NFC) films showed exceptional robustness under high pressures and moisture environments cycles. The optical band gap of this NFC film was found to be at 4.5 eV, with a corresponding HOMO energy position at 3.81 eV with respect to Ag. Comparing this value with *ab-initio* calculations for the electronic gap we obtain a rough estimate for the exciton binding energy of $E_{ex} \approx 800 meV$. Raman measurements as function of hydrostatic pressure revealed an irreversible modification of the NFC above 1.2 GPa attributed to a structural modification of the crystallites. In addition, peak force AFM measurements in controlled moisture conditions demonstrate that the elastic properties of the nanofibrillated cellulose can be reversibly controlled, showing an exceptional stability and reproducibility with deviations of less than 5% as function of moisture.

ASSOCIATED CONTENT

**Supporting Information**. Ab-initio calculations and additional characterization data are available in Supporting Information.

AUTHOR INFORMATION

[*] **Corresponding Author:** Claudia D. Simão; **claudia.simao@icn.cat**

[‡] **Present address:** Aalto University, Department of Applied Physics, P.O. Box 15100, Aalto, Finland

**Author Contributions**

The manuscript was written through contributions of all authors. All authors have given approval to the final version of the manuscript.

ACKNOWLEDGMENT

M.R.W. gratefully acknowledges the Marie Curie Fellowship (IEF) HeatProNano (Grant No. 628197). M.K. acknowledges the German Research Society (DFG) for a postdoctoral fellowship. This work was supported by the Spanish MINECO project TAPHOR (grant MAT-2012-31392). The authors acknowledge the nanometric techniques unit at CCiT-UB.

ABBREVIATIONS

NFC – nanofibrillated cellulose; BLS - Brillouin light scattering; QNM AFM – quantitative nanomechanical atomic force microscopy